\documentclass{elsart}
\usepackage[xdvi,dvipdf]{color}
\usepackage{amsmath,amssymb,epsfig}

\newcommand{\assign}{:=}

\newcommand{\mathi}{\mathrm{i}}
\newcommand{\tmmathbf}[1]{\ensuremath{\boldsymbol{#1}}}
\newcommand{\tmop}[1]{\ensuremath{\operatorname{#1}}}
\newcommand{\tmem}[1]{{\em #1\/}}
\newcommand{\um}{-}
\newcommand{\upm}{\pm}

\begin{document}
\begin{frontmatter}

\title{The twisted arc model for chiral molecules}
\author{Johannes Trost},
\author{Klaus Hornberger}
\address{Arnold Sommerfeld Center for Theoretical Physics, 
Ludwig-Maximilians-Universit{\"a}t M{\"u}nchen, Theresienstra{\ss}e 37, 
80333 Munich, Germany}

\begin{abstract}
We present a simple model for chiral molecules which yields the frequency-dependent multipole-multipole polarizabilities
required for calculating the dispersive van der Waals constants 
in a simple, closed, and consistent form.
The model consists of a single effective charge confined to a
one-dimensional wire formed by two circular arcs, which are twisted with respect to each other by an angle characterizing the chirality. The resulting 
polarizabilities show a simple functional dependence on the model parameters, and they serve to mimic the chiral properties of small molecules, such as H$_2$S$_2$, in a reasonably realistic way.
\end{abstract}

\end{frontmatter}

\section{Introduction}

Wire models {\cite{tinoco:onedimhelix,nolte:rotpoweronedim}} are popular to
describe molecular configurations where electrons are allowed to move almost
freely along quasi one-dimensional structures, as is the case e.g. with the
dislocated $\pi$-electrons in organic chain molecules
{\cite{balazs:twistedcircle,leuliette:helicoidalmodel,kondru:twistedwire}}.
Studies comparing these models with ab-initio electronic configuration
calculations {\cite{nolte:rotpoweronedim,power:largering}} show that, in spite
of their simplicity, these models succeed in reproducing many qualitative
features and yield even quantitatively reasonable results.

In the present article we discuss a simple wire model which aims at describing
small chiral molecules, and in particular the dependence of the chiral nature
on their electronic susceptibilities. The latter determine important molecular
properties such as the optical activity and the dispersion forces acting
between molecules. While the chiroptical properties have already been subject
of several publications
{\cite{tinoco:onedimhelix,nolte:rotpoweronedim,balazs:twistedcircle,leuliette:helicoidalmodel,kondru:twistedwire}},
the present article is focused on the dispersion forces.

In general, it is a difficult task to calculate the dispersive interaction
potential strengths between chiral molecules (with few or no symmetries) and
other polarizable particles if one has to go beyond the bulk van-der-Waals
force $U = - C_6 / r^6$. This is the case, e.g. if one needs to assess the
relative strength of chirality-discriminative part of the interaction versus
the non-discriminative ones. The reason is that multipole-multipole
polarizabilities are required on the whole imaginary frequency spectrum for
calculating the higher order dispersion constants which are linear
combinations of integrals {\cite{rijks:vanDerWaalsCoeff}}
\begin{equation}
  \int^{\infty}_0 \alpha^{\ell_A, \ell'_A}_{m_A, m'_A} \left( \mathi \omega
  \right) \alpha^{\ell_B, \ell'_B}_{m_B, m'_B} \left( \mathi \omega \right) d
  \omega, \label{polarIntegrals}
\end{equation}
where $\alpha^{\ell_A, \ell'_A}_{m_A, m'_A}$ and $\alpha^{\ell_B,
\ell'_B}_{m_B, m'_B}$ denote the polarizabilities for molecules A and B,
respectively, in spherical tensor notation, see Eq. (\ref{def:multipolePolar})
below.

The van-der-Waals dispersion constant $C_6$ depends on the
electric-dipole-electric-dipole polarizabilities of both molecules ($\ell_A =
\ell'_A = \ell_B = \ell'_B = 1$). While it does not discriminate between the
left- and right-handed configurations of an oriented chiral molecule, higher
order dispersion interactions differ in general for left and right
configurations. Often, a rotational average over all orientations of the
chiral molecule cancels this left-right distinction in the cross sections.
However, consider a coherent superposition of the left- and right-handed
configuration states {\cite{cinaHarris}}, which will be subject to decoherence
due to collisions with the atoms of a surrounding gas. In this case, the
decoherence process is determined by collisions of spherical gas atoms B off
{\tmem{the oriented}} chiral molecule A. Here, the lowest order discriminative
interaction with B ($\ell_B = \ell'_B = 1$) is given by the
electric-dipole-electric-quadrupole polarizability
{\cite{chiu:dispersionForces}}, $\ell_A = 1$ and $\ell'_A = 2$. The resulting
dispersive interaction depends on the distance $r$ between the molecules as
$r^{- 7}$ {\cite{stone:intermolecular}}. We will focus on this dominant
discriminative interaction below, motivated by the above-mentioned question of
collisional decoherence. We note that there might be additional forces of
comparable strength which do not discriminate left and right, and that only
non-retarded forces are considered in the following. For a discussion of
retarded interaction between chiral molecules, see
{\cite{mavroyannis,jenkins:dispersionbetweenChiralMolecules,craig:chiralDiscrimination}}.

The main aim of the present model is therefore to provide a minimalist, while
consistent and trustworthy model for the bulk electronic properties of chiral
molecules. In particular, we give explicit and closed results for the rotatory
power, the electric-dipole-electric-dipole polarizability, and the
electric-dipole-electric-quadrupole polarizability, both at imaginary
frequencies. The simplicity of the model even admits the calculation of
further higher multipole-multipole polarizabilities with small effort.

\begin{figure}[h]
  \resizebox{12cm}{!}{\epsfig{file=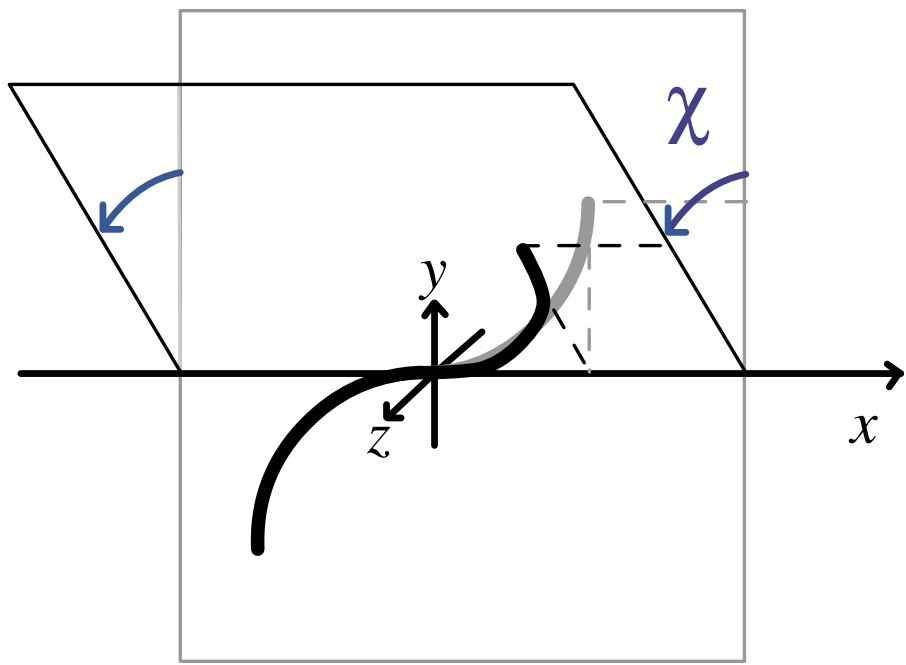}}
  \caption{The twisted arc model is given by a wire of length $L$ in the form
  of two arcs formed by quarter circles. They can be twisted with respect to
  each other by an angle $\chi$. This way a three-dimensional chiral structure
  is formed, except for $\chi = 0$ and $\chi = \pi$, when the wire is in its
  planar {\tmem{trans}}- and its
  {\tmem{cis}}-configuration.\label{fig:wire3d}}
\end{figure}

The model is specified by two parameters, the length $L$ determining the bulk
polarizability and the twist angle $\chi$, which corresponds to the dihedral
angle in a molecule, see Fig.~\ref{fig:wire3d}. By choosing the parameters to
represent the chiral dihydrogen disulfide, H$_2$S$_2$, we obtain reasonable
values for its various electronic properties, suggesting that the model may
serve as a good approximation for a typical small chiral molecule. Yet, the
aim here is of course not to reproduce precise numbers, but to have an easily
accessible, consistent and plausible description of the dependence of the
electronic susceptibilities on the molecular parameters, and in particular on
the dihedral angle.

The structure of the article is as follows. Sect.~\ref{sec:model} presents the
model and its quantization based on the canonical description of the
one-dimensional dynamics. The explicit forms of the most important Cartesian
operators (position, momentum, magnetic dipole, and electric quadrupole
moment) are then derived in Sect.~\ref{sec:cartesian}, together with closed
expressions for their matrix elements in the energy eigenbasis. Based on these
results the rotatory strength is calculated in Sect.~\ref{sec:rotatory}, as a
function of the chiral angle. Sect.~\ref{sec:alpha} contains the calculation
and discussion of the electric-dipole-electric-dipole and
electric-dipole-electric-quadrupole polarizability for imaginary frequencies.
An example for interaction strengths derived from the polarizabilities is
discussed in Sect.~\ref{sec:potential} and we present our conclusions in
Sect.~\ref{sec:conclusions}.

\section{The twisted arc model\label{sec:model}}

\begin{figure}[h]
  \resizebox{12cm}{!}{\epsfig{file=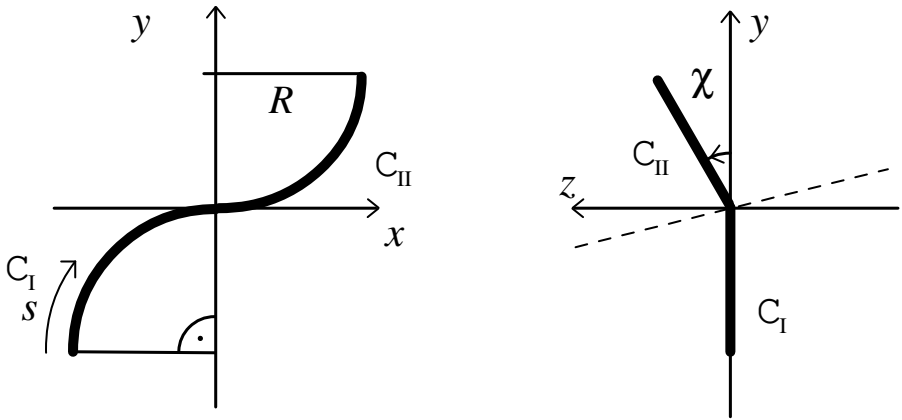}}\label{fig:wiredetails}
  \caption{Projections of the twisted wire model onto the $x y$-plane (left)
  and the $y z$-plane (right). The arcs $\mathcal{C}_{\text{I}}$ and
  $\mathcal{C}_{\text{II}}$ are quarter circles with a radius of $R = L /
  \pi$. $\mathcal{C}_{\text{I}}$ lies in the $x y$-plane as indicated and the
  twist angle $\chi$ is defined to be positive (negative) for
  $\mathcal{C}_{\text{II}}$ pointing into the positive (negative)
  $z$-direction. The origin of the parameterization coordinate $s$ is taken to
  be the loose end of $\mathcal{C}_{\text{I}}$. The $C_2$ symmetry axis, drawn
  as a dashed line, lies in the $yz$-pane.}
\end{figure}

Our model aims at describing the excitation properties of a chiral quantum
system in the easiest possible way. It is formed by two connected circular
arcs, $\mathcal{C}_{\text{I}}$ and $\mathcal{C}_{\text{II}}$, each described
by the common radius $R$ and an angle of 90$^{\circ}$. The total length of the
wire is thus $L = \pi R_{}$. We put the origin of the coordinate system at the
junction of the arcs such that $\mathcal{C}_{\text{I}}$ lies in the $x
y$-plane. For $\chi = 0$ also $\mathcal{C}_{\text{II}}$ lies in the $x
y$-plane ({\tmem{trans}}-configuration), while for $\chi \neq 0$ it is turned
around the $x$-axis, see Fig.~\ref{fig:wiredetails}. We call $\chi$ the twist
angle, and take it as the angle between the osculating plane of
$\mathcal{C}_{\text{II}}$ and the $y$-axis \ ($- \pi < \chi \leqslant \pi$).
For $\chi = \pi$ the wire is in its {\tmem{cis}}-configuration, and for $\chi
\neq 0, \pi$ the angles $\chi$ and $- \chi$ correspond to configurations with
opposite chirality. Alternatively one could choose the dihedral angle $\phi =
\pi - \chi$.

A charged particle of mass $M$ and charge $q$ is confined to move freely along
the wire defined by the two arcs. It is constrained by an infinite potential
step at both ends, but there is no force in the $x$-direction at the joint of
$\mathcal{C}_{\text{I}}$ and $\mathcal{C}_{\text{II}}$.

The model shows $C_2$ symmetry. The symmetry axis (dashed line in
Fig.~\ref{fig:wiredetails}) lies in the $y z$-plane and includes the origin.
It has an angle of $- \chi / 2$ with respect to the $z$-axis.

We note that an alternative construction, suggested in
{\cite{kondru:twistedwire}}, would be to take (at least) three straight wire
parts joint together at angles of 90$^{\circ}$. We prefer the present
cornerless structure, since it consists only of two equivalent legs which
mirror the symmetry of the wave function, and admit simple closed expressions
for the relevant matrix elements, see below. For a critical examination of the
thin wire model with corners see {\cite{power:onedimwire}}. Another widely
used model for chiral molecules is the helical thin wire model
{\cite{sears:nuclearMagneticShielding}}, which however does not allow the
straightforward modelling of the dihedral angle of H$_2$S$_2$.

\subsection{Quantization of the twisted arc\label{sec:wavefunction}}

In order to quantize the motion let us first consider the canonical
description of the classical motion {\cite{tinoco:onedimhelix}}. The natural
generalized coordinate is the length of the wire $s$ ($0 \leqslant s \leqslant
L$). Noting $R = L / \pi$ we find from Fig.~\ref{fig:wiredetails} its relation
to the Cartesian coordinates.
\begin{eqnarray}
  x & = & - \frac{L}{\pi} \cos \left( \frac{\pi s}{L} \right), \nonumber\\
  y & = & \left\{ \begin{array}{ll}
    \frac{L}{\pi}  \left[ \sin \left( \frac{\pi s}{L} \right) - 1 \right] &
    \text{for \ $0 \leqslant s \leqslant \frac{L}{2}$},\\
    \frac{L}{\pi} \cos \chi \left[ 1 - \sin \left( \frac{\pi s}{L} \right)
    \right] & \text{for \ $\text{$\frac{L}{2} \leqslant s \leqslant L$}$},
  \end{array} \right. \nonumber\\
  z & = & \left\{ \begin{array}{ll}
    0 & \text{ for \ $0 \leqslant s \leqslant \frac{L}{2}$},\\
    \frac{L}{\pi} \sin \chi \left[ 1 - \sin \left( \frac{\pi s}{L} \right)
    \right] & \text{ for \ $\text{$\frac{L}{2} \leqslant s \leqslant L$}$} .
  \end{array} \right.  \label{eq:coordinates}
\end{eqnarray}
As can be easily checked, this parameterization guarantees that the kinetic
energy $T$ is given by the generalized velocity
\begin{eqnarray}
  T & = &  \frac{M}{2}  \left( \dot{x}^2 + \dot{y}^2 + \dot{z}^2 \right) =
  \frac{M}{2}  \dot{s}^2 . 
\end{eqnarray}
In the absence of a (magnetic) potential the Lagrange function equals the
kinetic energy, $\mathcal{L}= T$. The conjugate momentum is defined as $p_s =
\partial \mathcal{L}/ \partial \dot{s} = M \dot{s_{_{\text{}}}}$, so that the
Hamilton function $H = T$ takes the form $H = p^2_s / \left( 2 M \right)$. The
Cartesian components of the momentum are thus obtained, via $p_x = M \dot{x} =
M (\partial x / \partial s) \dot{s}$, as
\begin{eqnarray}
  p_x & = & \sin \left( \frac{\pi s}{L} \right) p_s, \nonumber\\
  p_y & = & \left\{ \begin{array}{lll}
    & \cos \left( \frac{\pi s}{L} \right) p_s, & \text{for \ $0 \leqslant s
    \leqslant \frac{L}{2}$},\\
    & - \cos \chi \cos \left( \frac{\pi s}{L} \right) p_s, & \text{for \
    $\text{$\frac{L}{2} \leqslant s \leqslant L$}$},
  \end{array} \right. \nonumber\\
  p_z & = & \left\{ \begin{array}{lll}
    & 0, & \text{for \ $0 \leqslant s \leqslant \frac{L}{2}$},\\
    & - \sin \chi \cos \left( \frac{\pi s}{L} \right) p_s, & \text{for \
    $\text{$\frac{L}{2} \leqslant s \leqslant L$}$},
  \end{array} \right.  \label{eq:canMomenta}
\end{eqnarray}
We turn to the quantum description by replacing $p_s$ by the differential
operator
\begin{equation}
  \mathsf{p}_s = \frac{\hbar}{\mathi}  \frac{\partial^{}}{\partial s} .
  \label{momentumOperator}
\end{equation}
acting in $L_2 \left( \left[ 0, L \right] \right)$. Operators will be
indicated by sans-serifs throughout. The Hamiltonian is now
\begin{equation}
  \mathsf{H} = - \frac{\hbar^2}{2 M}  \frac{\partial^2}{\partial s^2} .
\end{equation}
Together with the boundary condition $\psi (0) = \psi (L) = 0$, it yields the
eigenfunctions of a particle in a one dimensional box with length $L$
\begin{equation}
  \psi_n (s) = \sqrt{\frac{2}{L}} \sin \left( \frac{n \pi s}{L} \right),
  \text{{\hspace*{\fill}}$n = 1, 2, 3, \ldots$} . \label{wavefunction1}
\end{equation}
The corresponding energies depend quadratically on the excitation number, $E_n
= n^2 E_1$, with the ground state energy given, for $M = m_e$, by
\begin{eqnarray}
  E_1 & = &  \frac{\pi^2}{2}  \left( \frac{L}{a_0} \right)^{- 2}
  E_{\text{hartree}} = \frac{35.06 \tmop{eV}}{(L / \text{{\AA}})^2} .
  \label{groundStateEnergy} 
\end{eqnarray}
Choosing $L$ as the sum of the binding lengths of H$_2$S$_2$, i.e. $L = 8.6
a_0$, yields a first excitation energy of $E_2 - E_1 = 3 E_1 = 5.4 \text{eV}$,
which is in good agreement with the result of ab initio calculations for the
excitation energy, $5.48 \text{eV}$, for the longest wavelength transition
{\cite{rauk:chiropticalProperties}}.

In the following, we will use atomic units (au), $\hbar = e_0 = m_e = 1$, with
$M = m_e$.

\section{The Cartesian operators\label{sec:cartesian}}

The functional dependence of the Cartesian operators on the coordinate $s$
will in general be different on $\mathcal{C}_{\text{I}}$ and
$\mathcal{C}_{\text{II}}$. It is therefore convenient to introduce for any
operator $\mathsf{A}$ the corresponding partial operators $\mathsf{A}^{\left(
\text{I} \right)}$ and $\mathsf{A}^{\left( \text{II} \right)}$ satisfying
\begin{equation}
  \mathsf{A} \left( s \right) = \Theta \left( \frac{L}{2} - s \right)
  \mathsf{A}^{\left( \text{I} \right)} \left( s \right) + \Theta \left( s -
  \frac{L}{2} \right) \mathsf{A}^{\left( \text{II} \right)} \left( s \right) .
\end{equation}
The matrix elements of $\mathsf{A}$ with respect to the energy eigenstates
(\ref{wavefunction1}) can then be calculated as
\begin{eqnarray}
  \left\langle m | \mathsf{A} | n \right\rangle & = & \int^{L / 2}_0 d s
  \psi_m^{\ast} (s) \mathsf{A}^{( \text{I})} \left( s \right) \psi_n (s) 
  \label{operatorSplit}\\
  &  & + \int_{L / 2}^{L_{}} d s \psi_m^{\ast} (s) \mathsf{A}^{( \text{II})}
  \left( s \right) \psi_n (s) . \nonumber
\end{eqnarray}
The $C_2$ symmetry of the wire model will allow to express matrix elements of
operator $\mathsf{A}$ by the restricted operator $\mathsf{A}^{\left( \text{I}
\right)}$ alone, see below.

\subsection{The position operator}

The Cartesian dipole operator is determined by the Cartesian components of the
position operator, $\overrightarrow{\mathsf{d} } = q
\overrightarrow{\mathsf{r}}$. From (\ref{eq:coordinates}) we obtain
immediately
\begin{eqnarray}
  \mathsf{x}^{\left( \text{I} \right)} & = & - \frac{L}{\pi} \cos \left(
  \frac{\pi s}{L} \right) = \mathsf{x}^{\left( \text{II} \right)}, 
  \label{coordinateOpsRepresentation}\\
  \mathsf{y}^{\left( \text{I} \right)} & = & \frac{L}{\pi}  \left[ \sin \left(
  \frac{\pi s}{L} \right) - 1 \right], \nonumber\\
  \mathsf{z}^{\left( \text{I} \right)} & = & 0, \nonumber\\
  \mathsf{y}^{\left( \text{II} \right)} & = &  \frac{L}{\pi} \cos \chi \left[
  1 - \sin \left( \frac{\pi s}{L} \right) \right] = - \cos \chi
  \mathsf{y}^{\left( \text{I} \right)}, \nonumber\\
  \mathsf{z}^{\left( \text{II} \right)} & = &  \frac{L}{\pi} \sin \chi \left[
  1 - \sin \left( \frac{\pi s}{L} \right) \right] . \nonumber
\end{eqnarray}
As an advantage of the present model, the matrix elements can be shown to be
simply interrelated,
\begin{eqnarray}
  \left\langle m \left| \mathsf{z}^{\left( \text{II} \right)} \right| n
  \right\rangle & = & \left( - \right)^{n + m + 1} \sin \chi \left\langle m
  \left| \mathsf{y}^{\left( \text{I} \right)} \right| n \right\rangle, 
  \label{propCoords}\\
  \left\langle m \left| \mathsf{y}^{\left( \text{II} \right)} \right| n
  \right\rangle & = & \left( - \right)^{n + m + 1} \cos \chi \left\langle m
  \left| \mathsf{y}^{\left( \text{I} \right)} \right| n \right\rangle .
  \nonumber
\end{eqnarray}
It follows that only a small fraction of the matrix elements needs to be
evaluated. For those we obtain
\begin{eqnarray}
  \left\langle m \left| \mathsf{x} \right| n \right\rangle & = & \left\langle
  m \left| \mathsf{x}^{\left( \text{I} \right)} \right| n \right\rangle +
  \left\langle m \left| \mathsf{x}^{\left( \text{II} \right)} \right| n
  \right\rangle  \label{coordMatrixEle}\\
  & = & \left\{ \begin{array}{ll}
    - \frac{L}{2 \pi}  \left( \delta_{n, m + 1} + \delta_{n, m - 1} \right) &
    \text{for $m > 1$,}\\
    - \frac{L}{2 \pi} \delta_{n, 2} & \text{for $m = 1$},
  \end{array} \right. \nonumber\\
  \left\langle m \left| \mathsf{y}^{\left( \text{I} \right)} \right| n
  \right\rangle & = & \frac{L}{\pi^2}  \frac{1}{D_{mn}^2 - 2 S_{mn} + 1} 
  \left[ \frac{2}{D_{mn}}  \left( n \left( 3 m^2 + n^2 - 1 \right) \gamma_n
  \sigma_m \right. \right. \nonumber\\
  &  & \left. - m \left( 3 n^2 + m^2 - 1 \right) \gamma_m \sigma_n - 4 mn
  \right] \nonumber\\
  & = & \left\langle n \left| \mathsf{y}^{\left( \text{I} \right)} \right| m
  \right\rangle \hspace{2em} \hspace{2em}  \text{for $m \neq n, n \pm 1$,}
  \nonumber\\
  \left\langle 1 \left| \mathsf{y}^{\left( \text{I} \right)} \right| n
  \right\rangle & = & \left\{ \begin{array}{ll}
    - \frac{2 L}{\pi^2}  \frac{2 D_{n 1} + \left( S_{n 1} + 1 \right)
    \gamma_n}{nD_{n 1} D_{n 2}}  & \text{for $n > 2$,}\\
    - \frac{L}{3 \pi^2}  & \text{for $n = 2$ .}
  \end{array} \right. \nonumber
\end{eqnarray}
Here we introduced abbreviations which will be used throughout the paper,
\begin{eqnarray}
  S_{mn} & \assign & m^2 + n^2, \nonumber\\
  D_{mn} & \assign & m^2 - n^2 .  \label{eq:SDdef}
\end{eqnarray}
Moreover, the following factors show up frequently:
\begin{eqnarray}
  &  & \gamma_m \assign \cos \left( \frac{m \pi}{2} \right) = \left\{
  \begin{array}{ll}
    1 & \text{for $m = 0 \tmop{mod} 4$},\\
    - 1 & \text{for $m = 2 \tmop{mod} 4$},\\
    0 & \text{otherwise} ;
  \end{array} \right.  \label{eq:cdef}\\
  &  & \sigma_m \assign \sin \left( \frac{m \pi}{2} \right) = \left\{
  \begin{array}{ll}
    1 & \text{for $m = 1 \tmop{mod} 4$},\\
    - 1 & \text{for $m = 3 \tmop{mod} 4$},\\
    0 & \text{otherwise.}
  \end{array} \right.  \label{eq:sdef}
\end{eqnarray}
The matrix elements of dipole operators in spherical form are calculated from
the basic matrix elements Eq. (\ref{coordMatrixEle})
\begin{eqnarray}
  \left\langle 1 | \mathsf{d}_0 | n \right\rangle & = & q \left\langle 1 |
  \mathsf{z}^{\left( \text{II} \right)} | n \right\rangle = \left( - \right)^n
  \sin \chi q \left\langle 1 \left| \mathsf{y}^{\left( \text{I} \right)}
  \right| n \right\rangle,  \label{sphericalDipol}\\
  \left\langle 1 | \mathsf{d}_{\pm 1} | n \right\rangle & = & \mp
  \frac{q}{\sqrt{2}} \left[ \left\langle 1 \left| \mathsf{x} \right| n
  \right\rangle \pm \mathi \left\langle 1 \left| \mathsf{y} \right| n
  \right\rangle \right] \nonumber\\
  & = & \mp \frac{q}{\sqrt{2}} \left[ \left\langle 1 \left| \mathsf{x}
  \right| n \right\rangle \pm \mathi \left( 1 - \left( - \right)^n \cos \chi
  \right) \left\langle 1 \left| \mathsf{y}^{\left( \text{I} \right)} \right| n
  \right\rangle \right] . \nonumber
\end{eqnarray}

\subsection{The momentum operator\label{sec:momentum}}

The Cartesian components of the momentum operator on the individual arcs can
be obtained for the corresponding classical expressions (\ref{eq:canMomenta}).
However, unlike the full operator $\mathsf{p}$, the hermiticity of
$\mathsf{p}^{( \text{I})}$ and $\mathsf{p}^{( \text{II})}$ is not guaranteed.

By replacing the canonical momentum $p^{}_s$ in the symmetrized version of
(\ref{eq:canMomenta}) by its corresponding operator (\ref{momentumOperator}),
and carrying out the derivatives as far as possible, we get the Cartesian
momentum operators
\begin{eqnarray}
  \mathsf{p}^{( \text{I})}_x & = & - \mathi \left[ \sin \left( \frac{\pi s}{L}
  \right)  \frac{\partial}{\partial s} + \frac{\pi}{2 L} \cos \left( \frac{\pi
  s}{L} \right) \right] = \mathsf{p}_x^{\left( \text{II} \right)}, 
  \label{momentumOperators}\\
  \mathsf{p}^{( \text{I})}_y & = & - \mathi \left[ \cos \left( \frac{\pi s}{L}
  \right)  \frac{\partial}{\partial s} - \frac{\pi}{2 L} \sin \left( \frac{\pi
  s}{L} \right) \right], \nonumber\\
  \mathsf{p}^{( \text{II})}_y & = & - \mathi \cos \chi \left[ - \cos \left(
  \frac{\pi s}{L} \right)  \frac{\partial}{\partial s} + \frac{\pi}{2 L} \sin
  \left( \frac{\pi s}{L} \right) \right] = - \cos \chi \mathsf{p}_y^{\left(
  \text{I} \right)}, \nonumber\\
  \mathsf{p}^{( \text{I})}_z & = & 0, \nonumber\\
  \mathsf{p}^{( \text{II})}_z & = & - \mathi \sin \chi \left[ - \cos \left(
  \frac{\pi s}{L} \right)  \frac{\partial}{\partial s} + \frac{\pi}{2 L} \sin
  \left( \frac{\pi s}{L} \right) \right] = - \sin \chi \mathsf{p}^{\left(
  \text{I} \right)}_y . \nonumber
\end{eqnarray}
The operators $\mathsf{p}^{( \text{I})}_y$, $\mathsf{p}^{( \text{II})}_y$,
$\mathsf{p}^{( \text{I})}_z$, and $\mathsf{p}^{( \text{II})}_z$ are hermitian,
while $\mathsf{p}^{\left( \text{I} \right)}_x$ and $\mathsf{p}^{\left(
\text{II} \right)}_x$, are individually not hermitian, but only their sum. For
example, $\mathsf{p}^{\left( \text{I} \right)}_x$ obeys the relation
\begin{equation}
  \left\langle n | \mathsf{p}_x^{\left( \text{I} \right)} | m \right\rangle =
  \left\langle m | \mathsf{p}^{\left( \text{I} \right)}_x | n
  \right\rangle^{\ast} - \frac{2}{L} \mathi s_n s_m .
\end{equation}
The second term on the right hand side is the boundary term of the partial
integration showing up if the differential operator is applied to the bra. It
does not necessarily vanish here, since the wave function may be finite at the
boundary point $s = L / 2$. A remedy would be to define a modified momentum
operator
\begin{equation}
  \mathsf{\tilde{p}}^{\left( \text{I} \right)}_x \assign \mathsf{p}^{\left(
  \text{I} \right)}_x + 2 \mathi \delta \left( s - \frac{L}{2} \right) .
  \label{momentumICorrected}
\end{equation}
This operator on $\mathcal{C}_{\text{I}}$ is now hermitian, $\langle m|
\widetilde{\mathsf{p}}^{\left( \text{I} \right)}_x |n \rangle = \langle n|
\widetilde{\mathsf{p}}^{\left( \text{I} \right)}_x |m \rangle^{\ast}$, since
the $\delta$-function cancels the contribution of the boundary term.
Similarly, the definition
\begin{equation}
  \mathsf{\tilde{p}}^{\left( \text{II} \right)}_x \assign \mathsf{p}^{\left(
  \text{II} \right)}_x - 2 \mathi \delta \left( s - \frac{L}{2} \right)
  \label{momentumIICorrected}
\end{equation}
ensures both the hermiticity on $\mathcal{C}_{\text{II}}$ and the relation
$\mathsf{p}_x = \widetilde{\mathsf{p}}_x^{\left( \text{I} \right)} +
\widetilde{\mathsf{p}}_x^{\left( \text{II} \right)}$. In general, it would be
therefore more convenient to express momentum matrix elements in terms of the
hermitian operators (\ref{momentumICorrected}) and
(\ref{momentumIICorrected}). However, below the momentum operator will occur
only as part of the magnetic dipole operator, where this modification is not
required, as discussed next.

\subsection{Magnetic dipole moments\label{sec:magnetic}}

The magnetic moment is proportional to the angular momentum operator
\begin{equation}
  \overrightarrow{\mathsf{m}} = q \overrightarrow{\mathsf{L}} = q
  \overrightarrow{\mathsf{r}} \times \overrightarrow{\mathsf{p}} = q
  \left(\begin{array}{c}
    \mathsf{y}  \mathsf{p}_z - \mathsf{z}  \mathsf{p}_y\\
    \mathsf{z}  \mathsf{p}_x - \mathsf{x}  \mathsf{p}_z\\
    \mathsf{x}  \mathsf{p}_y - \mathsf{y}  \mathsf{p}_x
  \end{array}\right),
\end{equation}
where we still use atomic units and allow for an effective charge $q =
Z_{\tmop{eff}} e_0$, which might be useful as a fitting parameter.

The magnetic dipole moment depends on the origin, and it is natural to choose
the position of the joint between the two arcs in the present model. It can be
easily seen that the non-hermiticity of the parts $\mathsf{p^{\left( \text{I}
\right)}_x}$ and $\mathsf{p}^{\left( \text{II} \right)}_x$ is then irrelevant
since the $x$-component of the momentum is multiplied by $y$ or $z$, which
vanish at $s = L / 2$ and thus suppress the hermiticity terms in Eqs.
(\ref{momentumICorrected}) and (\ref{momentumIICorrected}). The ability to
evaluate the magnetic moment directly, without having to resort to hermitian
corrections, is another advantage of the present model.

Again one can derive helpful inter-relations of the matrix elements,
\begin{eqnarray}
  \left\langle m | \left( \mathsf{x}  \mathsf{p}_y - \mathsf{y} 
  \mathsf{p}_x)^{\left( \text{II} \right)} | n \right\rangle \right. & = &
  \left( - \right)^{n + m} \cos \chi \left\langle m | \left( \mathsf{x} ^{}
  \mathsf{p}_y - \mathsf{y}^{} \mathsf{p}_x \right)^{\left( \text{I} \right)}
  | n \right\rangle  \label{eq:mrelations}\\
  \left\langle m | \mathsf{z}  \mathsf{p}_x - \mathsf{x}  \mathsf{p}_z | n
  \right\rangle & = & \left\langle m | \left( \mathsf{z}  \mathsf{p}_x -
  \mathsf{x}  \mathsf{p}_z \right)^{\left( \text{II} \right)} | n
  \right\rangle \nonumber\\
  & = & \left( - \right)^{n + m + 1} \sin \chi \left\langle m | \left(
  \mathsf{x}  \mathsf{p}_y - \mathsf{y}  \mathsf{p}_x)^{\left( \text{I}
  \right)} | n \right\rangle, \right. \nonumber\\
  \left\langle m | \mathsf{y}  \mathsf{p}_z - \mathsf{z}  \mathsf{p}_y | n
  \right\rangle & = & \left\langle m | \left( \mathsf{y}  \mathsf{p}_z -
  \mathsf{z} \mathsf{p}_y \right)^{\left( \text{II} \right)} | n \right\rangle
  = 0. \nonumber
\end{eqnarray}
The last relation follows immediately from the proportionality of $y$ and $z$
components of the operators and the vanishing of the $z$ components on
$\mathcal{C}_{\text{I}}$. It follows that a single integral remains to be
calculated.
\begin{eqnarray}
  \mathi \left\langle m | \left( \mathsf{x}  \mathsf{p}_y - \mathsf{y} 
  \mathsf{p}_x \right)^{\left( \text{I} \right)} | n \right\rangle & = &
  \left\{ \begin{array}{ll}
    0 & \text{for $n = m$,}\\
    - \frac{4 \left( 4 - \pi \right) m \left( m + 1 \right) - \pi}{8 \left( 2
    m + 1 \right) \pi} & \text{for $n = m + 1$.}
  \end{array} \right. 
\end{eqnarray}
For $n > m + 1$ we get
\begin{eqnarray}
  \mathi \left\langle m | \left( \mathsf{x}  \mathsf{p}_y - \mathsf{y} 
  \mathsf{p}_x \right)^{\left( \text{I} \right)} | n \right\rangle & = & -
  \left[ 2 mn \left( \Delta^2_{mn} - 1 \right) \left( \Sigma^2_{mn} - 1
  \right)  \right.\\
  &  & + 2 mn \left( 2 S_{mn} - 1 \right) \gamma_m \gamma_n \nonumber\\
  &  & \left. + \left( 2 S_{mn}^2 - D^2_{mn} - S_{mn} \right) \sigma_m
  \sigma_n  \right] \nonumber\\
  &  & / \left[ \pi D_{mn} \left( \Delta^2_{mn} - 1 \right) \left(
  \Sigma^2_{mn} - 1 \right) \right] . \nonumber
\end{eqnarray}
Here we added the abbreviations
\begin{eqnarray}
  \Sigma_{mn} & \assign & m + n \\
  \Delta_{mn} & \assign & m - n \nonumber
\end{eqnarray}
to those defined in (\ref{eq:SDdef})-(\ref{eq:sdef}). The hermiticity of the
part $\mathsf{m}^{\left( \text{I} \right)}_z$ is now evident, $\langle n|
\mathsf{m}^{\left( \text{I} \right)}_z |m \rangle = \langle m|
\mathsf{m}^{\left( \text{I} \right)}_z |n \rangle^{\ast}$, implying with
(\ref{eq:mrelations}) that all the components of $\overrightarrow{\mathsf{m}}$
are hermitian.

\subsection{Electric quadrupole moments\label{sec:electricQuad}}

The quadrupole moments are given by a quadratic combination of the position
operators. In Cartesian coordinates, $\vec{r} = \left( x, y, z \right)$, we
have
\begin{equation}
  \mathsf{\Theta}_{ij} = q \left( 3 r_i r_k - \delta_{ik}  \vec{r}^2  \right)
  . \label{def:quadrupoleMom}
\end{equation}
Again, a variety of proportionality relations serves to reduce the
calculational effort considerably,
\begin{eqnarray}
  \left\langle n | \mathsf{y}^{\left( \text{II} \right)2} | m \right\rangle^{}
  & = & \cos^2 \chi \left( - 1 \right)^{n + m}  \left\langle n |
  \mathsf{y}^{\left( \text{I} \right)2} | m \right\rangle^{}, \\
  \left\langle n | \mathsf{z}^2 | m \right\rangle & = & \left\langle n |
  \mathsf{z}^{\left( \text{II} \right)2} | m \right\rangle \nonumber\\
  & = & \sin^2 \chi \left( - 1 \right)^{n + m} \left\langle n |
  \mathsf{y}^{\left( \text{I} \right)2} | m \right\rangle^{}, \nonumber\\
  \left\langle n | \mathsf{x}^{\left( \text{II} \right)}  \mathsf{y}^{\left(
  \text{II} \right)} | m \right\rangle & = & \cos \chi \left( - 1 \right)^{n +
  m}  \left\langle n | \mathsf{x}^{\left( \text{I} \right)} \mathsf{y}^{\left(
  \text{I} \right)} | m \right\rangle, \nonumber\\
  \left\langle n | \mathsf{xz} | m \right\rangle & = & \left\langle n |
  \mathsf{x}^{\left( \text{II} \right)}  \mathsf{z}^{\left( \text{II} \right)}
  | m \right\rangle \nonumber\\
  & = & \sin \chi \left( - 1 \right)^{n + m} \left\langle n |
  \mathsf{x}^{\left( \text{I} \right)}  \mathsf{y}^{\left( \text{I} \right)} |
  m \right\rangle, \nonumber\\
  \left\langle n | \mathsf{yz} | m \right\rangle & = & \left\langle n |
  \mathsf{y}^{\left( \text{II} \right)} \mathsf{z}^{\left( \text{II} \right)}
  | m \right\rangle \nonumber\\
  & = & \sin \chi \cos \chi \left( - 1 \right)^{n + m} \left\langle n |
  \mathsf{y}^{\left( \text{I} \right)2} | m \right\rangle . \nonumber
\end{eqnarray}
It follows that three integrals need to be evaluated,
\begin{eqnarray}
  \left\langle n | \mathsf{x}^2 | m \right\rangle & = & \frac{L^2}{2 \pi^2}
  \left( \delta_{mn} + \frac{1}{2} \delta_{m, n + 2} + \frac{1}{2} \delta_{m,
  n - 2} \right),  \label{eq:x2me}\\
  \left\langle n | \mathsf{y}^{\left( \text{I} \right)2} | m \right\rangle & =
  & \frac{4 L^2}{\pi^2}  \left[ 2 nm \left( \Delta_{mn}^2 - 4 \right) D_{mn} 
  \label{eq:y2me} \right.\\
  &  & + 3 n \left( 5 m^4 + n^4 + 10 m^2 n^2 - 15 m^2 - 5 n^2 + 4 \right)
  \gamma_n \sigma_m \nonumber\\
  &  & \left. - 3 m \left( 5 n^4 + m^4 + 10 m^2 n^2 - 15 n^2 - 5 m^2 + 4
  \right) \gamma_m \sigma_n \right] \nonumber\\
  &  & \left[ D_{mn} \left( \Delta_{mn}^2 - 4 \right) \left( \Delta_{mn}^2 -
  1 \right) \left( \Sigma_{mn}^2 - 4 \right) \left( \Sigma_{mn}^2 - 1 \right)
  \right]^{- 1}, \nonumber
\end{eqnarray}
\begin{eqnarray}
  \left\langle n | \mathsf{x}^{\left( \text{I} \right)} \mathsf{y}^{\left(
  \text{I} \right)} | m \right\rangle & = & \frac{2 L^2}{\pi^3} \left[ - 2 nm
  \left( \Delta_{mn}^2 - 1 \right) \left( \Sigma_{mn}^2 - 1 \right) 
  \label{eq:xyme} \right.\\
  &  & + 6 mn \left( 2 S_{mn} - 5 \right) \gamma_n \gamma_m \nonumber\\
  &  & \left. + 3 \left( 2 S^2_{mn} - D_{mn} - 5 S_{mn} + 4 \right) \sigma_n
  \sigma_m \right] \nonumber\\
  &  & \times \left[ \left( \Delta^2_{mn} - 4 \right) \left( \Delta_{mn}^2 -
  1 \right) \left( \Sigma_{mn}^2 - 4 \right) \left( \Sigma_{mn}^2 - 1 \right)
  \right]^{- 1} . \nonumber
\end{eqnarray}
The last two expressions, (\ref{eq:y2me}) and (\ref{eq:xyme}), are valid for
those combinations of values $n$ and $m$ for which the denominators do not
vanish. We omit the general results for $m - n = 1, 2$ since the matrix
elements will be needed only for $m = 1$ below. In this case they take the
form
\begin{eqnarray}
  \left\langle n | \mathsf{y}^{\left( \text{I} \right)2} | 1 \right\rangle & =
  & \frac{4 L^2}{\pi^3 nD_{n, 2}} \left( 2 - \frac{3 \left( n^2 + 6 \right)
  \gamma_n}{D_{n, 1} D_{n, 3}} \right), \text{ \ \ \ \ \ \ for $n > 3$}, \\
  \left\langle 2 | \mathsf{y}^{\left( \text{I} \right)2} | 1 \right\rangle & =
  & \frac{2 L^2}{15 \pi^3}, \nonumber\\
  \left\langle 3 | \mathsf{y}^{\left( \text{I} \right)2} | 1 \right\rangle & =
  & \frac{2 L^2}{\pi^3} \left( \frac{4}{15} - \frac{\pi}{16} \right),
  \nonumber\\
  \left\langle n | \mathsf{x}^{\left( \text{I} \right)} \mathsf{y}^{\left(
  \text{I} \right)} | 1 \right\rangle & = & \frac{2 L^2}{\pi^3}  \frac{2
  nD_{n, 2} + 3 S_{n, 1} \sigma_n}{D_{n, 1} D_{n, 2} D_{n, 3}}, \text{ \ \ \
  for $n > 3$}, \nonumber\\
  \left\langle 2 | \mathsf{x}^{\left( \text{I} \right)} \mathsf{y}^{\left(
  \text{I} \right)} | 1 \right\rangle & = & \frac{4 L^2}{\pi^3} \left(
  \frac{\pi}{16} - \frac{2}{15} \right), \nonumber\\
  \left\langle 3 | \mathsf{x}^{\left( \text{I} \right)} \mathsf{y}^{\left(
  \text{I} \right)} | 1 \right\rangle & = & \frac{7 L^2}{30 \pi^3} . \nonumber
\end{eqnarray}
Using these results one gets quite compact expressions for the quadrupole
moments, which display a simple dependence on the twist angle $\chi$. Here we
note the matrix elements of the spherical quadrupole operators,
$\mathsf{Q}_{2, \mu}$, with magnetic quantum number $\mu$. They are required
for the calculation of the electric-dipole-electric-quadrupole polarizability
in tensorial form, $\alpha^{1, 2}_{k, k'} \left( \omega \right)$, see below
\begin{eqnarray}
  \left\langle 1 | \mathsf{Q}_{2, 0} | n \right\rangle & \assign & \frac{q}{2}
  \left\langle 1 | 2 \mathsf{z}^2 - \mathsf{x}^2 - \mathsf{y}^2 | n
  \right\rangle  \label{sphericalQuad}\\
  & = & \um \frac{q}{2}  \left\langle 1 | \mathsf{x}^{\left( \text{I}
  \right)2} | n \right\rangle - \frac{q}{2}  \left[ \left( - 1 \right)^n
  \left( 3 \sin^2 \chi - 1 \right) + 1 \right]  \left\langle 1 |
  \mathsf{y}^{\left( \text{I} \right)2} | n \right\rangle, \nonumber\\
  &  &  \nonumber\\
  \left\langle 1 | \mathsf{Q}_{2, \pm 1} | n \right\rangle & \assign & \mp q
  \sqrt{\frac{3}{2} } \left\langle 1 | \mathsf{z}  \mathsf{x} \pm \mathi
  \mathsf{z}  \mathsf{y} | n \right\rangle \nonumber\\
  & = & \pm q \sqrt{\frac{3}{2}} \sin \chi \left( - 1 \right)^n \left(
  \left\langle 1 | \mathsf{x}^{\left( \text{I} \right)} \mathsf{y}^{\left(
  \text{I} \right)} | n \right\rangle \pm \mathi \cos \chi \left\langle 1 |
  \mathsf{y}^{\left( \text{I} \right)2} | n \right\rangle \right), \nonumber\\
  \left\langle 1 | \mathsf{Q}_{2, \pm 2} | n \right\rangle & \assign & q
  \sqrt{\frac{3}{8}}  \left\langle 1 | \mathsf{x}^2 \pm 2 \mathi \mathsf{x} 
  \mathsf{y} - \mathsf{y}^2 | n \right\rangle \nonumber\\
  & = & q \sqrt{\frac{3}{8}}  \left\{ \left\langle 1 | \mathsf{x}^{\left(
  \text{I} \right)2} | n \right\rangle \right. \nonumber\\
  &  & \pm 2 \mathi \left[ 1 - \left( - 1 \right)^n \cos \chi \right]
  \left\langle 1 | \mathsf{x}^{\left( \text{I} \right)} \mathsf{y}^{\left(
  \text{I} \right)} | n \right\rangle \nonumber\\
  &  & \left. - \left[ 1 - \left( - 1 \right)^n \cos^2 \chi \right] 
  \left\langle 1 | \mathsf{y}^{\left( \text{I} \right)2} | n \right\rangle
  \right\} . \nonumber
\end{eqnarray}

\section{The rotatory strength\label{sec:rotatory}}

We proceed with the discussion of the rotatory strength, the most prominent
property of chiral molecules. We do so to demonstrate the ability of the
twisted arc model to display electromagnetic properties of real, chiral
molecules. The latter are usually optically active, and their chiroptical
properties depend on the optical rotatory tensor {\cite{craig:molecularQED}},
\begin{equation}
  R^{n 1}_{ij} = \tmop{Im} \left\{ \left\langle 1 | \mathsf{d}_i | n
  \right\rangle \left\langle n | \mathsf{m}_j | 1 \right\rangle \right\} .
  \text{\label{eq:opticalRotTensor}}
\end{equation}
The trace of the optical rotatory tensor yields the rotatory strength,
\begin{equation}
  R^{n 1} = R^{n 1}_{xx} + R^{n 1}_{yy} + R^{n 1}_{zz} = \tmop{Im} \left\{
  \left\langle 1 | \overrightarrow{\mathsf{d}} | n \right\rangle \cdot
  \left\langle n | \overrightarrow{\mathsf{m}} | 1 \right\rangle \right\} .
\end{equation}
It determines the optical rotation angle for orientationally averaged
molecules. Recognizing an energy-resolution of the identity operator one finds
that the sum over all states must vanish,
\begin{equation}
  \sum_{n = 1}^{\infty} R^{n 1} = \tmop{Im} \sum^{\infty}_{n = 1} \left\langle
  1 | \overrightarrow{\mathsf{d}} | n \right\rangle \cdot \left\langle n |
  \overrightarrow{\mathsf{m}} | 1 \right\rangle = \tmop{Im} \left\langle 1 |
  \overrightarrow{\mathsf{d}} \cdot \overrightarrow{\mathsf{m}} | 1
  \right\rangle = 0, \label{eq:sumrule}
\end{equation}
because the expectation value of the hermitian operator
$\overrightarrow{\mathsf{d}} \cdot \overrightarrow{\mathsf{m}}$ must be real.
We note also that the rotatory strength does not depend on the choice of
origin. The most convenient choice of origin is therefore the joint of the
arcs, since hermiticity corrections of the momentum operator are then not
required, as discussed above. Note that an object independent of the origin is
obtained by amending the rotatory tensor (\ref{eq:opticalRotTensor}) with
additional terms depending on the quadrupole-dipole tensor, which however
cancel out after rotational averaging {\cite{buckingham:orientedMolecules}}.

From the above formulas for the matrix elements of
$\overrightarrow{\mathsf{m}}$ and $\overrightarrow{\mathsf{d}}$ one obtains
\begin{eqnarray}
  R^{n 1} & = & Z^2_{\tmop{eff}} L \frac{4}{\pi^3} \sin \chi \left( - 1
  \right)^n \\
  &  & \times \left[ \frac{4}{D_{n, 1} D_{n, 2}} + \frac{2 \left( n^2 + 2
  \right) \gamma_n}{D_{n, 1}^2 D_{n, 2}} + \frac{2 \left( n^2 + 5 \right)
  \sigma_n}{nD_{n, 1} D_{n, 2}^2} \right], \text{ \ \ \ \ \ \ \ \ \ for $n >
  2$,} \nonumber\\
  R^{21} & = & Z^2_{\tmop{eff}} L \frac{4}{3 \pi^3} \sin \chi \left(
  \frac{2}{3} - \frac{3 \pi}{16} \right) . 
\end{eqnarray}
As one expects, the rotatory strength is antisymmetric with respect to the
twist angle, and the sum rule (\ref{eq:sumrule}) for the rotatory strength is
fulfilled.

Reasonable parameters for a small chiral molecule such as for $\text{H}_2
\text{S}_2$ are $L = 8.6$ and $Z^2_{\tmop{eff}} = 4.5$ (see
Sect.~\ref{sec:results}). With these values one finds that $R^{21}$ for $\chi
= \frac{\pi}{2}$ is $0.388$ au, corresponding to about $180 \times 10^{- 40}$
esu, which is larger by a factor of 5 to 10 compared with theoretical values
for $\text{H}_2 \text{S}_2$
{\cite{rauk:chiropticalProperties,pericou:circularDichroismH2S2}}. This is
mainly due to cancellation effects generated by the two lowest lying excited
states of $\text{H}_2 \text{S}_2$, which are nearly degenerated for dihedral
angles around $\pi / 2$. They have rotatory strengths which are similar in
absolute values but opposite in sign {\cite{rauk:chiropticalProperties}}.

Having found reasonable choices for the parameters $L$ and $Z_{\tmop{eff}}$,
we now proceed to evaluate the frequency dependent rotatory power $G' \left(
\omega \right)$, a further quantity which permits a comparison with the
literature values of the real molecule H$_2$S$_2$. The rotatory power is
expressed by the electric-dipole-magnetic-dipole polarizability
{\cite{buckingham:forces}},
\begin{equation}
  G_{ij}' \left( \omega \right) = - 2 \omega \sum_{n = 2}^{\infty}
  \frac{R_{ij}^{n 1}}{\left( E_n - E_1 \right)^2 - \omega^2} .
\end{equation}
The average over all orientations of $G_{ij}'$ is given by the trace
\begin{equation}
  G' \left( \omega \right) = \sum_{i = x, y, z} G'_{ii} \left( \omega \right)
  = - 2 \omega \sum_{n = 2}^{\infty} \frac{R^{n 1}}{\left( E_n - E_1 \right)^2
  - \omega^2}, \label{eq:rotPower1}
\end{equation}
which can be expanded for small frequencies (compared to the excitation gap).
For $\chi = \pi / 2$ we find
\begin{equation}
  G_{\pi / 2}' \left( \omega \right) Z^{- 2}_{\tmop{eff}} L^{- 3} = - 1.142
  \cdot 10^{- 4} \omega - 1.61 \cdot 10^{- 5} \omega^3 + O \left( \omega^5
  \right) . \label{eq:rotPower2}
\end{equation}
Here, the frequencies are in units of $E_1 / \hbar$. The polarizability, given
in atomic units, scales as $Z^2_{\tmop{eff}} L^3$, where a factor
$Z^2_{\tmop{eff}} L$ is contributed by the dimensions of the rotatory tensor
and a factor $L^2$ by the decrease of the excitation energy with increasing
length scale.

The specific rotation angle $\phi$ per dm can be calculated as (see e.g.
{\cite{stone:intermolecular}})
\begin{equation}
  \phi \left( \omega \right) = - 187.5^{\circ} \tmop{dm}^{- 1} \sin \chi
  \frac{\eta}{\left( \tmop{mol} / \tmop{dl} \right)}  \frac{\hbar \omega}{E_1}
  G_{\pi / 2}' \left( \frac{\hbar \omega}{E_1} \right), \label{eq:rotAngle1}
\end{equation}
where $\eta$ is the concentration of chiral molecules modeled by twisted
wires. The validity of (\ref{eq:rotAngle1}) is restricted to frequencies well
below the first excitation energy. Inserting the frequency for the sodium
D-line ($E_{\tmop{Na}} \approx 1.17 E_1$ for $L = 8.6 a_0$), with the above
choice of $L = 8.6 a_0$ and $Z^2_{\tmop{eff}} = 4.5$, one gets
\begin{equation}
  \phi_{\text{Na-D}} = - 103^{\circ} \tmop{dm}^{- 1} \sin \chi
  \frac{\eta}{\left( \tmop{mol} / \tmop{dl} \right)} . \label{eq:rotAngle2}
\end{equation}
The specific rotation angle of H$_2$S$_2$ has a somewhat different functional
dependency on $\chi$ due to abovementioned cancellation effects, but its value
is about the same order of magnitude as in CI calculations
{\cite{pericou:circularDichroismH2S2}}.

Note that the optical rotatory tensor (\ref{eq:opticalRotTensor}) appears in
the discriminatory part of the dispersion interaction between {\tmem{two}}
chiral molecules (here without rotational averaging)
{\cite{craig:molecularQED}},
\begin{equation}
  E_{\tmop{ch} - \tmop{ch}} = \frac{2}{R^6}  \left( \delta_{ik} - 3 \hat{R}_i
  \hat{R}_k \right) \left( \delta_{jl} - 3 \hat{R}_j \hat{R}_l \right)
  \sum_{n, m} \frac{R^{n 1}_{ij} R^{m 1}_{kl}}{E_{n 1} + E_{m 1}} .
  \label{chiralChiralInteraction}
\end{equation}
We will not evaluate the expression further since we are mainly concerned with
interactions between a chiral molecule and achiral atoms.

\section{Electric multipole-multipole polarizabilities\label{sec:alpha}}

We proceed to calculate the electric multipole-multipole polarizability tensor
required for evaluating the dispersion interaction coefficients, $C_6$ and
$C_7$. With this application in mind, the results will not be presented in
full generality, but will be confined to the ground state polarizability at
imaginary frequencies.

\subsection{General form of the electric-dipole-electric-dipole polarizability
tensor}

Apart from $\alpha_{xx}$, all elements of the polarizability tensor
$\tmmathbf{\alpha}$ depend on the twist angle $\chi$. We start by discussing
this dependence in more detail, noting that the elements of the polarization
tensor of the state $m$ are given as {\cite{buckingham:forces}}
\begin{eqnarray}
  \alpha^{(m)}_{i j} (\omega, \chi) & = & \sum^{\infty}_{n \neq m} \left[
  \frac{\left\langle m | \mathsf{d}_i (\chi) | n \right\rangle \left\langle n
  | \mathsf{d}_j (\chi) | m \right\rangle}{E_n - E_m - \omega}  \right.\\
  &  & \left. \hspace*{\fill} \hspace*{\fill} \phantom{\left[ \right.} +
  \frac{\left\langle m | \mathsf{d}_j (\chi) | n \right\rangle \left\langle n
  | \mathsf{d}_i (\chi) | m \right\rangle}{E_n - E_m + \omega} \right],
  \nonumber
\end{eqnarray}
where $i, j \in \{x, y, z\}$.

Due to the time-reversal invariance the tensor elements are real at imaginary
frequencies, see e.g. {\cite{landau:QM}}, implying that the tensor is a
symmetric matrix and can hence be written as
\begin{equation}
  \alpha^{\left( m \right)}_{i j} (\mathi \omega, \chi) = 2 \sum^{\infty}_{n
  \neq m} \frac{(E_n - E_m)}{(E_n - E_m)^2 + \omega^2}  \left\langle m |
  \mathsf{d}_i (\chi) | n \right\rangle \left\langle n | \mathsf{d}_j (\chi) |
  m \right\rangle . \label{defPolarTensor}
\end{equation}
For the special case $m = 1$ and inserting $\overrightarrow{\mathsf{d}} =
Z_{\text{eff}} \overrightarrow{\mathsf{r}}$ we get finally
\begin{equation}
  \alpha^{\left( 1 \right)}_{i j} (\mathi \omega, \chi) = \frac{4 L^2
  Z^2_{\text{eff}}}{\pi^2} \sum^{\infty}_{n = 2} \frac{(n^2 - 1) \left\langle
  1 | \mathsf{r}_i (\chi) | n \right\rangle \left\langle n | \mathsf{r}_j
  (\chi) | 1 \right\rangle}{(n^2 - 1)^2 + \omega^2},
\end{equation}
where we absorbed a factor $\pi^2 / 2 L^2$ into the frequency, thus measuring
the frequency $\omega$ again in units of $E_1 / \hbar$, see
(\ref{groundStateEnergy}). The index indicating the reference state $m = 1$
will be suppressed from now on.

We find that, due to the interrelations of the position matrix elements noted
above, the frequency dependence of the spherical tensor is determined by only
three functions,
\begin{eqnarray}
  g_{\text{e}} \left( \omega \right) & = & \frac{4}{\pi^2 L^2} \sum_{n = 2, 4,
  \ldots}  \frac{n^2 - 1}{\left( n^2 - 1 \right)^2 + \omega^2}  \left\langle n
  | \mathsf{y}^{\left( \text{I} \right)} | 1 \right\rangle^2, \hspace{2em} 
  \label{def:gA}\\
  g_{\text{o}} \left( \omega \right) & = & \frac{4}{\pi^2 L^2}  \sum_{n = 3,
  5, \ldots}  \frac{n^2 - 1}{\left( n^2 - 1 \right)^2 + \omega^2} 
  \left\langle n | \mathsf{y}^{\left( \text{I} \right)} | 1 \right\rangle^2,
  \nonumber\\
  a \left( \omega \right) & = & \frac{1}{\pi^4 }  \frac{3}{9 + \omega^2} .
  \nonumber
\end{eqnarray}
With their help the structure of $\tmmathbf{\alpha}$ at imaginary frequencies
is readily specified,
\begin{eqnarray}
  &  & \tmmathbf{\alpha}(\mathi \omega, \chi) = Z^2_{\text{eff}} L^4 
  \label{generalAlpha}\\
  &  &  \nonumber\\
  &  & \times \left(\begin{array}{ccc}
    a \left( \omega \right) & \frac{2}{3 \pi}  \left( 1 + \text{c}_{\chi}
    \right) a \left( \omega \right) & \frac{2}{3 \pi} s_{\chi} a \left( \omega
    \right)\\
    \frac{2}{3 \pi}  \left( 1 + c_{\chi} \right) a \left( \omega \right) & g_+
    \left( \omega \right) \left( 1 + c_{\chi}^2 \right) + 2 g_- \left( \omega
    \right) c_{\chi} & \left( g_+ \left( \omega \right) + g_- \left( \omega
    \right) c_{\chi} \right) s_{\chi}\\
    \frac{1}{3 \pi} s_{\chi} a \left( \omega \right) & \left( g_+ \left(
    \omega \right) + g_- \left( \omega \right) c_{\chi} \right) s_{\chi} & g_+
    \left( \omega \right) s_{\chi}^2 
  \end{array}\right) \nonumber
\end{eqnarray}
Here we used the abbreviations $s_{\chi} \assign \sin \chi$ and $c_{\chi}
\assign \cos \chi$ and $g_{\pm} \left( \omega \right) \assign g_{\text{e}}
\left( \omega \right) \upm g_{\text{o}} \left( \omega \right)$. The sums in
(\ref{def:gA}) converge rapidly as the terms are of order $O \left( n^{- 6}
\right)$ for large $n$.

The dependence on the twist angle is particularly simple in
(\ref{generalAlpha}) due to the proportionality relations (\ref{propCoords}).
One observes that only the off-diagonal elements in the third row and the
third column change their sign when switching to opposite handedness, $\chi
\rightarrow - \chi$. Such a change of sign in $\chi$ is equivalent to a
reflection at the $x y$-plane, $z \rightarrow - z$. After an additional
rotation by $\pi$ around the $z$-axis a full parity operation, $\vec{r}
\rightarrow - \vec{r}$, is obtained. Hence, one gets the tensor of the $-
\chi$ configuration after rotating the polarizability tensor of the $+ \chi$
configuration, which illustrates the well known fact that all dispersive
interactions solely derived from the electric-dipole-electric-dipole
polarizability tensor do not discriminate left- and right-handed enantiomers.

\subsection{Choice of parameters\label{sec:results}}

In the following we present numerical results for a specific choice of
parameters, which are adapted to mimic the literature values for H$_2$S$_2$,
e.g. {\cite{rauk:chiropticalProperties,pericou:circularDichroismH2S2}}. First
we fix the length parameter $L$ by the sum of the binding lengths of the
molecule which is found to be 8.6 $a_0$. The twist angle in the wire model is
chosen to be $\chi = \pm \frac{\pi}{2}$ which is a good approximation of the
dihedral angle in the molecule.

\begin{figure}[h]
  \resizebox{12cm}{!}{\epsfig{file=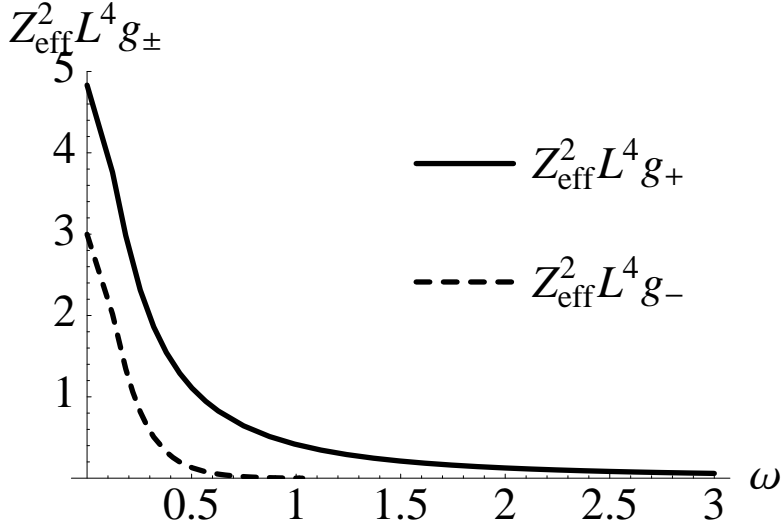}}
  \caption{Frequency dependence of the functions $Z_{\text{eff}}^2 L^4 g_+
  (\omega)$ (solid line) and $Z^2_{\text{eff}} L^4 g_- \left( \omega \right)$
  (dashed line) [atomic units; $L = 8.6$; $Z^2_{\tmop{eff}} = 4.5$]. The
  frequency is given in terms of the ground state energy $E_1 / \hbar$, see
  (\ref{groundStateEnergy}).\label{fig:gPlusMinus}}
\end{figure}

To fix the effective charge $Z_{\text{eff}}$ we consider the static (i.e.,
$\omega = 0$) values for the electric-dipole-electric-dipole polarizability
$g_+ \left( 0 \right) = 1.963 \times 10^{- 4}$ au, $g_- \left( 0 \right) =
1.218 \times 10^{- 4}$ au, and $a \left( 0 \right) = 3.42 \times 10^{- 3}$ au.
The values of $g_+$ and $g_-$ as function of $\omega$ are depicted in
Fig.~\ref{fig:gPlusMinus}. A comparison with the Lorentzian $Z_{\text{eff}}^2
L^4 a \left( \omega \right)$ from (\ref{def:gA}), which has a width of 3 and a
strength of 84.2, shows that the latter dominates the polarizability tensor.

The rotational average of the static polarizability for $\chi = \pm \pi / 2$
is then given by
\begin{eqnarray}
  L^4 Z^2_{\text{eff}}  \bar{\alpha} \left( 0 \right) & = & \frac{L^4
  Z^2_{\text{eff}}}{3} \left( \alpha_{xx} \left( 0 \right) + \alpha_{yy}
  \left( 0 \right) + \alpha_{zz} \left( 0 \right) \right) \\
  & \approx & L^4 Z^2_{\text{eff}} 1.271 \times 10^{- 3}  \text{au} .
  \nonumber
\end{eqnarray}
Here the contribution of $\alpha_{xx} \left( 0 \right)$ dominates the average
static polarizability. A comparison with the literature value
{\cite{rauk:chiropticalProperties}} for $\text{H}_2 \text{S}_2$ of about 31 au
results in $Z^2_{\text{eff}} = 4.5$ ($Z_{\text{eff}} \approx 2.12$).

\subsection{Electric-dipole-electric-quadrupole polarizability\label{sec:A}}

The dominant discriminative dispersion interaction between an oriented chiral
and an achiral atom or molecule depends on the
electric-dipole-electric-quadrupole polarizability tensor
{\cite{chiu:dispersionForces}}. In addition to the dipole matrix elements, it
is determined by the matrix elements quadratic in the coordinate operators,
$\left\langle n \left| \mathsf{r}_i  \mathsf{r}_j \right| m \right\rangle$,
discussed in Sect.~\ref{sec:electricQuad}.

For the calculation of dispersion constants it is preferable to consider
polarizability tensors for imaginary frequencies in spherical representation
\begin{eqnarray}
  \alpha^{\ell, \ell'}_{k, k'} \left( \mathi \omega \right) & = & \frac{1}{2
  \pi}  \sqrt{\left( 2 \ell + 1 \right) \left( 2 \ell' + 1 \right)} 
  \label{def:multipolePolar}\\
  &  & \times \sum_{n = 2}^{\infty} \frac{E_n - E_1}{\left( E_n - E_1
  \right)^2 - \left( \mathi \omega \right)^2}  \left\langle 1 |
  \mathsf{Q}_{\ell, k} | n \right\rangle \left\langle n | \mathsf{Q}_{\ell',
  k'} | 1 \right\rangle, \nonumber
\end{eqnarray}
with the spherical multipole functions
\begin{equation}
  Q_{\ell, k} \left( \vec{r} \right) = Z_{\text{eff}} r^{\ell}  \sqrt{\frac{4
  \pi}{2 \ell + 1}} Y_{\ell, k} \left( \theta, \varphi \right) .
\end{equation}
For the electric-dipole-electric-quadrupole polarizability the multipole
operators have ranks $\ell = 1$ and $\ell' = 2$,
\begin{equation}
  \alpha^{1, 2}_{k, k'} \left( \mathi \omega \right) = \frac{\sqrt{15}}{2 \pi}
  \sum_{n = 2}^{\infty} \frac{E_n - E_1}{\left( E_n - E_1 \right)^2 +
  \omega^2}  \left\langle 1 | \mathsf{d}_k | n \right\rangle \left\langle n |
  \mathsf{Q}_{2, k'} | 1 \right\rangle . \label{dipoleQuadrupoleTensor}
\end{equation}
The required matrix elements can be found in (\ref{sphericalDipol}) and
(\ref{sphericalQuad}). Obviously, the polarizability $\alpha_{k, k'}^{1, 2}$
transforms under parity operation like a product of three coordinates,
$\mathsf{r}_i \mathsf{r}_j \mathsf{r}_l$, which results in an overall negative
sign. Thus, the dispersive interaction derived from it discriminates
left-handed and right-handed molecules. Finally, it should be noted that the
electric-dipole-electric-dipole polarizability $\alpha^{1, 1}$ does not depend
on the choice of the origin, while the electric-quadrupole-electric-dipole
polarizability $\alpha^{2, 1}$ does {\cite{stone:intermolecular}}.

\section{Potential strengths\label{sec:potential}}

As an illustration for the use of the polarizability tensors, let us evaluate
the interaction potential between a helium atom in ground state and the
twisted arc. Its calculation requires the electric-dipole-electric-dipole
polarizability of helium. Since helium is spherically symmetric only a single
element of the spherical polarizability tensor does not vanish, namely
$\alpha_{0, 0}^{1, 1}$. For our purposes it suffices to approximate the helium
polarizability by
\begin{equation}
  \alpha_{0, 0}^{1, 1} \left( \mathi \omega \right) =
  \frac{g}{\omega^2_{\tmop{He}} + \omega^2},
\end{equation}
with $g = 2$ and $\omega_{\text{He}} \approx 1.33$ (atomic units), which is a
good approximation for frequencies well below the excitation energy
{\cite{chan:hePolar}}.

After evaluating the integrals (\ref{polarIntegrals}) the dispersion
potentials can be calculated following the theory in
{\cite{osinga:vanDerWaals}} by disregarding retardation effects. They assume
the form
\begin{equation}
  U_6 \left( \vec{r} \right) = - \frac{C_6 \left( \hat{r} \right)}{r^6}
  \hspace{2em} \text{and} \hspace{2em} U_7 \left( \vec{r} \right) = -
  \frac{C_7 \left( \hat{r} \right)}{r^7}
\end{equation}
with $\vec{r}$ the distance vector between helium atom and the center of mass,
and $\hat{r} = \vec{r} / r$. For simplicity, we take the center of mass to lie
in the origin. The potential strength $C_6 \left( \hat{r} \right)$ is
calculated using the spherical dipole-dipole polarizability tensor,
$\alpha^{1, 1}_{kk'}$, of the twisted wire model which is readily obtained
from the Cartesian polarizability (\ref{generalAlpha}). The calculation of
$C_7 \left( \hat{r} \right)$ uses the dipole-quadrupole tensor
(\ref{dipoleQuadrupoleTensor}).

To compare the interaction potentials we consider the surface of equality,
\begin{equation}
  r_{\text{eq}} \left( \hat{r} \right) = \frac{\left| C_7 \left( \hat{r}
  \right) \right|}{\left| C_6 \left( \hat{r} \right) \right|} .
\end{equation}
Thus, at distances $r = \xi r_{\tmop{eq}}$ the strengths of the potentials are
related by $\left| U_6 \right| = \xi |U_7 |$. Since the potential strengths
are linear combinations of the polarizabilities $r_{\text{eq}} \left( \hat{r}
\right)$ is proportional to $L$. At the same time, the dependence on
$Z_{\text{eff}}$ cancels out.

\begin{figure}[h]
  \resizebox{12cm}{!}{\epsfig{file=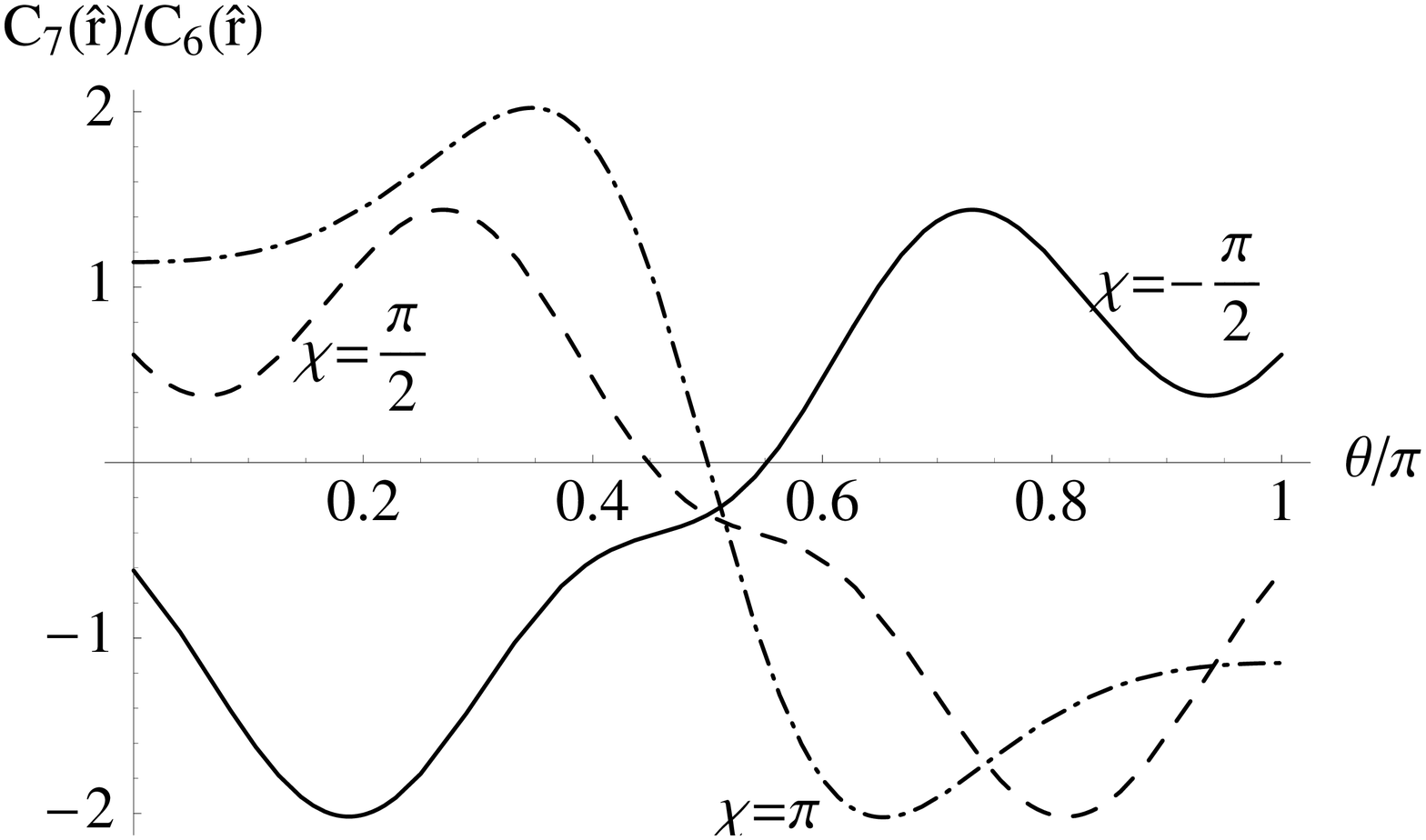}}
  \caption{\label{fig:relativeStrength}The relative strength of the
  discriminative potential $C_7 \left( \hat{r} \right)$ compared to the
  dipole-dipole potential $C_6 \left( \hat{r} \right)$ in Bohr radii as a
  function of the polar angle $\theta$ (see text). The azimuthal angle is
  chosen to be $\varphi = \frac{\pi}{2}$ and the values of the three different
  twist angles $\chi$ are indicated in the figure.}
\end{figure}

Figure~\ref{fig:relativeStrength} shows $C_7 / C_6$ for twist angles $\chi =
\pi$ and $\chi = \pm \pi / 2$ at $L = 8.6 a_0$. \ Here, the direction of the
helium atom, $\hat{r} = \left( \theta, \varphi \right)$, is specified with
respect to the symmetry axis of the twisted wire (the dashed line in
Fig.~\ref{fig:wiredetails}). The azimuthal angle (with respect to the
$x$-axis) is chosen to be $\varphi = \pi / 2$. Negative values of $C_7 / C_6$
indicate that $U_7 \left( \vec{r} \right)$ is repulsive. As can be seen, the
achiral cis-configuration $\chi = \pi$ is anti-symmetric with respect to
$\theta = \pi / 2$ (dash-dotted line). This is the case for all $\varphi$, and
it is a consequence of the additional reflection symmetry for a planar
configuration. For the other achiral situation, $\chi = 0$ (the
trans-configuration), the potential strength $C_7$ vanishes identically, $C_7
= 0$. For all pairs of left- and right-handed configurations, $\chi \neq 0,
\pi$, the surfaces of equality cannot be transformed into each other by a
proper rotation. Thus, unlike $U_6$, the $U_7$ interaction does distinguish
between right- and left-handed form, the difference depending on both the
distance $r = \xi r_{\tmop{eq}}$, and the orientation $\hat{r}$.

\[  \]
\section{Conclusions}\label{sec:conclusions}

We described a minimalist model for chiral molecules, which, in spite of its
simplicity, admits a physically plausible and consistent description of their
chiral properties. As a big advantage of this twisted arc model, the
evaluation of higher order electric moments can be reduced to a small number
of simple functions given in closed form. This way the functional dependence
of the multipole polarizabilities on the frequency and on the model parameters
shows up transparently. The derived chiral properties, such as the rotatory
strength and the dispersive interaction potentials, thus display a simple,
while physically consistent dependence on the model parameters. This was
demonstrated, specifically for the dihedral angle, by evaluating the
chirality-distinguishing part of the dispersive interaction with a polarizable
atom for molecular parameters adopted to the dihydrogen disulfide molecule.
The comparison with the dominant bulk interaction thus permits to assess to
what degree right- and left-handed, oriented molecules can be distinguished by
the dispersion interaction.

This work was supported by the Emmy Noether program of the DFG.

\end{document}